\documentclass[prb,aps,twocolumn,superscriptaddress]{revtex4-1}

\usepackage{braket}
\usepackage{amssymb,amsmath}
\usepackage{xspace}
\usepackage{graphicx}
\usepackage{hyperref}
\usepackage[all]{hypcap}
\usepackage{floatrow}
\usepackage[caption=false]{subfig}
\DeclareMathOperator{\Tr}{Tr}

\newcommand{\<}{\langle}

\newcommand{\up}{\uparrow}
\newcommand{\dn}{\downarrow}
\renewcommand{\>}{\rangle}
\renewcommand{\(}{\left(}
\renewcommand{\)}{\right)}
\renewcommand{\[}{\left[}
\renewcommand{\]}{\right]}

\newcommand{\Eq}[1]{Eq.~\ref{#1}}
\newcommand{\Fig}[1]{Fig.~\ref{#1}}
\newcommand{\Sec}[1]{Sec.~\ref{#1}}
\newcommand{\Ref}[1]{Ref.~\onlinecite{#1}}

\newcommand*{\Renyi}{R\'enyi\xspace}

\begin{document}

\title{Partial breakdown of quantum thermalization in a Hubbard-like model}

\author{James R. Garrison}
\affiliation{Department of Physics, University of California, Santa Barbara, California 93106, USA}
\affiliation{Joint Quantum Institute and Joint Center for Quantum Information and Computer Science, National Institute of Standards and Technology and University of Maryland, College Park, Maryland 20742, USA.}
\author{Ryan V. Mishmash}
\affiliation{Department of Physics and Institute for Quantum Information and Matter, California Institute of Technology, Pasadena, CA 91125, USA}
\affiliation{Walter Burke Institute for Theoretical Physics, California Institute of Technology, Pasadena, CA 91125, USA}
\author{Matthew P. A. Fisher}
\affiliation{Department of Physics, University of California, Santa Barbara, California 93106, USA}

\begin{abstract}
We study the possible breakdown of quantum thermalization in a model of itinerant electrons on a one-dimensional chain without disorder, with both spin and charge degrees of freedom.  The eigenstates of this model exhibit peculiar properties in the entanglement entropy, the apparent scaling of which is modified from a ``volume law'' to an ``area law'' after performing a partial, site-wise measurement on the system.  These properties and others suggest that this model realizes a new, non-thermal phase of matter, known as a quantum disentangled liquid (QDL).  The putative existence of this phase has striking implications for the foundations of quantum statistical mechanics.
\end{abstract}

\maketitle

\section{Introduction} \label{sec:intro}

In recent years, physicists have made great strides toward better understanding the equilibration and thermalization of isolated, many-body quantum systems.  Already, two distinct phases are well known: there exist systems that thermalize completely, such that for an arbitrary initial pure state any sufficiently small subregion will eventually approach the Gibbs ensemble; and, by contrast, there are systems that exhibit many-body localization (MBL) due to a strong disorder potential, failing to thermalize at any time despite weak interactions.

In a system that does approach thermal equilibrium, energy, particles, and other conserved quantities propagate throughout such that the system acts as its own bath.  After equilibration, any sufficiently small subregion will approximate the thermal density matrix (Gibbs ensemble), and all observables within any small subregion will match their values in the canonical ensemble.  One of the most important steps toward understanding quantum thermalization occurred in the early 1990s, when Deutsch and Srednicki independently proposed that thermalization, when it occurs, happens at the level of each individual eigenstate of finite energy density \cite{Deutsch1991-eth, Srednicki1994-thermalization}.  This result is generally known as the ``eigenstate thermalization hypothesis'' (ETH) \cite{Srednicki1996, Srednicki1999-eth, Rigol2008-eth, DePalma2015-eth}.  Within the framework of ETH, the ultimate fate of a system can be determined by examining the properties of its finite energy density eigenstates, without needing to consider the detailed quantum dynamics.  In fact, a single eigenstate of such a system directly reproduces the thermal ensemble in an arbitrarily-large subregion $A$ as long as the ratio of the subsystem to system size $V_A / V$ approaches zero as $V \rightarrow \infty$ \cite{Garrison2015x-eth}.  Also, the von Neumann entanglement entropy $S_A$ within the subsystem will match the thermal entropy, scaling as the volume of the subsystem, $S_A \sim L_A^d$ as long as $V_A < V/2$.  This is in contrast with typical ground states, which scale as an ``area law,'' $S_A \sim L_A^{d-1}$ \cite{Bombelli1986-entropy, Srednicki1993-area-law}.  In fact, the mechanism of thermalization can be thought of as the spreading of entanglement: each subsystem becomes maximally entangled with the remainder of the system over time, to the extent allowed by conservation laws (such as the conservation of energy).

One well-known counterexample to quantum thermalization is given by integrable systems, such as the one-dimensional Hubbard model (which is solvable via Bethe ansatz \cite{Lieb1968-hubbard-1d, Essler2005-book}).  Integrable systems typically have an infinite number of conserved quantities, which scales with total system size.  While these conserved quantities can (in some cases) permit analytic solution, they also prohibit full thermalization.  Instead of relaxing to a Gibbs ensemble, these systems relax to a ``generalized Gibbs ensemble,'' which takes into account the additional conservation laws \cite{Rigol2007-gge, Calabrese2011-gge}.  It is perhaps surprising that integrable systems exhibit a ``weak'' form of ETH: nearly all states appear locally thermal, but there exist rare, non-thermal states which are responsible for the breakdown of thermalization \cite{Biroli2010-rare}.  While integrable systems are interesting examples of systems that do not fully thermalize, they are tuned to special (solvable) points in parameter space, and are therefore non-generic.

As mentioned above, there also exist \emph{non-integrable}, interacting many-body quantum systems which do not thermalize and instead exhibit many-body localization (MBL) \cite{Anderson1958-localization, Gornyi2005-localization, Basko2006-mbl, Oganesyan2007-localization, Pal2010-mbl-transition, Bauer2013-mbl, Imbrie2014x-mbl}.  These systems have a strong disorder potential and sufficiently weak interactions, and are characterized by zero DC conductivity and partial memory of the initial state at all times.  Remarkably, the strong disorder potential leads to an ``emergent'' integrability, with resulting local integrals of motion \cite{Serbyn2013-iom, Huse2014-iom, Imbrie2014x-mbl}.  Due to these additional conservation laws, eigenstates in an energy window $\Delta E$ are very different from one another, and there are no avoided crossings between neighboring eigenstates when a parameter in the Hamiltonian is varied.  This results in energy-level spacings obeying Poisson statistics, in contrast to Wigner-Dyson statistics in a thermal system \cite{Oganesyan2007-localization}.  Additionally, the finite energy density eigenstates of these systems typically exhibit area law scaling of entanglement entropy ($S_A \sim L_A^{d-1}$), similar to a quantum ground state but in contrast to thermal systems (which instead exhibit volume law scaling) \cite{Bauer2013-mbl}.  Recently, MBL has been demonstrated in experiments on cold atomic gases \cite{DeMarco2013-mbl-expt, Bloch2015x-mbl-expt, Choi2016x-2d-mbl} and trapped ions \cite{Smith2015x-mbl-expt}, thus putting a vast amount of theoretical and numerical work on an experimental footing.  Conceptually, the existence of many-body localization provides an example of a system with a complete breakdown of thermalization, thus calling into question the general validity of quantum statistical mechanics.

It is tempting to wonder whether a phase of matter could exist between the extremes of full thermalization and MBL in a generic (i.e.\ non-integrable), isolated, many-body quantum system.  Many-body localization can be viewed as a situation in which infinitely massive (i.e.\ stationary) particles cause a classical disorder potential.  Given this line of thinking, one might be tempted to ask: what if the particles are quantum mechanical, allowed to move with a very large (but finite) mass?  Could a phase similar to MBL exist in such a translationally invariant, fully quantum mechanical system?  Guided by this question, \Ref{Grover2014-qdl} proposed a new phase of matter in multi-component liquids with two species of indistinguishable particles with a large mass ratio.  This phase, the ``quantum disentangled liquid'' (QDL), is characterized by heavy particles which are fully thermalized, but light particles which have not thermalized independently of the heavy particles.  Other work has also considered the possibility that thermalization can break down in translationally invariant systems\cite{Kagan1974, Kagan1984, Carleo2012-localization, Schiulaz2013x-mbl-glass, Schiulaz2015-mbl, Hickey2014-mbl, DeRoeck2014-thermal-mbl, DeRoeck2014-ti-mbl, DeRoeck2014x-ti-mbl-review, Yao2014x-quasi-mbl, Papic2015x-ti-mbl, VanHorssen2015-ti-mbl, He2015x-hubbard-mbl, Kim2016-ti-localization}.

In addition to proposing the QDL phase, \Ref{Grover2014-qdl} provided a qualitative diagnostic to identify eigenstates in the phase.  This diagnostic can be phrased in terms of entanglement entropy after a partial measurement.  As mentioned above, systems that fully thermalize exhibit volume law scaling for their entanglement entropy, while many-body localized systems exhibit an area law.  The QDL phase, like the fully ergodic phase, is characterized by eigenstates that are in an overall volume law for the entanglement entropy.  However, after a partial measurement of the locations of the heavy particles, the resulting wavefunction of the light particles is instead characterized by an area law in the QDL phase.  This suggests that the light particles are ``localized'' by the heavy particles, and is in contrast to a fully ergodic system, where the entanglement entropy of the light particles would scale as a volume law even after the measurement of the heavy particles' positions.  The proposed phase is called a ``quantum disentangled liquid'' because a partial measurement results in a ``disentangled'' wavefunction, a smoking gun for the breakdown of full thermalization.

The diagnostic given in \Ref{Grover2014-qdl} is very general and can be applied to any multi-component system.  In this paper, we will focus on 1D itinerant electron models with two species of fermions (spin-up and spin-down) on a lattice, specifically the Hubbard model with an additional nearest-neighbor repulsion term, which breaks integrability.  Instead of considering light and heavy particles, we will consider to what degree the spin and charge degrees of freedom thermalize independently from one another.  Overall, our results demonstrate the clear existence of a QDL regime at all system sizes that are accessible by full exact diagonalization.

The paper is organized as follows.  In \Sec{sec:diagnostic}, we introduce the QDL diagnostic for a lattice system with spin-half particles.  In \Sec{sec:model}, we introduce the Hubbard model with an additional nearest-neighbor repulsion term, which forms the basis for the remainder of the paper.  \Sec{sec:numerical} describes in detail our method for performing numerical exact diagonalization on this model.  In \Sec{sec:doublon_results}, we study each eigenstate's average doublon occupation, an observable which appears to violate ETH in the large-$U$ limit of the non-integrable model. In \Sec{sec:entropy_results}, we study the entanglement entropy properties of eigenstates, both before and after a partial measurement on each site.  In \Sec{sec:discussion}, we discuss implications for cold atom experiment and for the foundations of quantum statistical mechanics.

\section{Entanglement entropy diagnostic} \label{sec:diagnostic}

In this section we review and expound the diagnostic introduced in \Ref{Grover2014-qdl} for identifying quantum disentangled eigenstates, which is applicable to multi-component quantum systems on a lattice or in the continuum.  While \Ref{Grover2014-qdl} focused on systems with mass-imbalanced particles, here we will instead consider lattice systems with two species of fermions (spin-up and spin-down), with both spin and charge degrees of freedom.  The single-site Hilbert space consists of empty, spin-up, spin-down, and doubly-occupied states, which are denoted by $\ket{-}$, $\ket{\up}$, $\ket{\dn}$, and $\ket{\up\dn}$ respectively.

Let us first review the standard formulation of entanglement entropy.  Given a pure state $\ket\psi$ and a spatial subregion $A$ of size $L_A^d$ (where $d$ is the number of dimensions), the reduced density matrix in region $A$ is given by $\rho_A(\ket\psi) = \Tr_{\overline{A}} \ket\psi \bra\psi$, where $\overline{A}$ is the spatial complement of region $A$.  The von Neumann entanglement entropy in subregion $A$ is then given by $S_A(\ket\psi) = - \Tr_A \[ \rho_A(\ket\psi) \ln \rho_A(\ket\psi) \]$.  In a thermal system this quantity scales extensively with the subsystem size ($S(\ket\psi) \sim L_A^d$), but in a many-body localized system it scales as the size of its boundary, $S(\ket\psi) \sim L_A^{d-1}$.  These two possibilities are commonly known as ``volume law'' and ``area law,'' respectively.  The scaling of the overall entanglement entropy thus provides insight into whether a system is localized or not \cite{Bauer2013-mbl}.

The goal of the QDL diagnostic is to identify volume law states in which spin and charge have not thermalized independently of each other, despite the degrees of freedom having entangled with one another.  Guided by this intuition, the diagnostic considers the entanglement entropy after a \emph{partial measurement}, e.g.\ of the spin on each site.  If performing the partial measurement transforms a state from a volume law to an area law state, then the remaining degrees of freedom in the wavefunction have not thermalized independently of the measured degrees of freedom.  The remainder of this section explains this diagnostic in detail.

Consider a finite energy density eigenstate $\ket\psi$ of a system with overall volume law scaling of the entanglement entropy ($S_A \sim L_A^d$).  Given $\ket\psi$, we can perform a partial projective (von Neumann) measurement to determine the spin on each site along the $z$-axis, which results in a collapsed wavefunction
\begin{equation}
\ket{\phi_{\{\sigma\}}} = \frac{P_{\{\sigma\}} \ket\psi}{\sqrt{\bra\psi P_{\{\sigma\}} \ket\psi}}
\end{equation}
corresponding to some overall spin configuration ${\{\sigma\}}$.  Here, $P_{\{\sigma\}}$ is the projector onto the subspace consistent with the measurement outcome $\{\sigma\}$, and the probability of outcome $\{\sigma\}$ is given by the Born rule: $\text{Prob}(\{\sigma\}) = \bra\psi P_{\{\sigma\}} \ket\psi$.  Note that in our notation $\ket{\phi_{\{\sigma\}}}$ has been rescaled to have unit norm.

The state $\ket{\phi_{\{\sigma\}}}$ after the spin measurement is a wavefunction in which only charge degrees of freedom remain.  If a site has spin $+\frac12$ or $-\frac12$ along the $z$-axis, the charge on that site is one; however, if a site has overall spin zero, then it is possible that the site has either charge 0 or charge 2.  The wavefunction $\ket{\phi_{\{\sigma\}}}$ is thus a partially-collapsed state in which sites with spin zero can be in a superposition of two different charge states.  As a concrete example, let us consider a wavefunction $\ket\psi$ on a system with length $L=4$ and $N_\up=N_\dn=2$.  Say, for instance, that a partial measurement of the spins along the $z$-axis gives $[0, -\frac12, 0, +\frac12]$.  Then the charge on sites 2 and 4 is known, but sites 1 and 3 can be in a superposition of charge 0 and 2.  The resulting wavefunction is thus $\ket{\phi_{\{\sigma\}}} = \alpha \( \ket{-} \otimes \ket{\dn} \otimes \ket{\up\dn} \otimes \ket{\up} \) + \beta \( \ket{\up\dn} \otimes \ket{\dn} \otimes \ket{-} \otimes \ket{\up} \)$, where the values $\alpha$ and $\beta$ can be calculated given full knowledge of the original state $\ket\psi$.

In order to quantify the remaining amount of entanglement in the partially-collapsed state $\ket{\phi_{\{\sigma\}}}$, we consider the scaling of its entanglement entropy.  Given a subsystem $A$ of size $L_A^d$ and a measurement outcome $\{\sigma\}$, the reduced density matrix in region $A$ is given by $\rho_A(\ket{\phi_{\{\sigma\}}}) = \Tr_{\overline{A}} \ket{\phi_{\{\sigma\}}} \bra{\phi_{\{\sigma\}}}$ and the entanglement entropy is $S_A(\ket{\phi_{\{\sigma\}}}) = - \Tr_A \[ \rho_A(\ket{\phi_{\{\sigma\}}}) \ln \rho_A(\ket{\phi_{\{\sigma\}}}) \]$.  By averaging over all possible measurement outcomes with their associated Born-rule probabilities, we can calculate the average post-measurement entanglement entropy,
\begin{equation}
S_A^{c/s} \equiv \sum_{\{\sigma\}} \text{Prob}(\{\sigma\}) \, S_A(\ket{\phi_{\{\sigma\}}}),
\end{equation}
where $c/s$ denotes the entropy of the resulting charge wavefunction after a measurement of the spin on each site.  It is instructive to consider the scaling of this entanglement entropy taken after the partial measurement.  In a fully ergodic system, it should scale as a volume law for any partial measurement which does not fully collapse the wavefunction.  If the post-measurement entanglement entropy instead scales as an area law, then the charge has not thermalized independently of the spin, and the system is non-ergodic.

It is also possible to consider a diagnostic which reverses the roles of spin and charge (i.e.\ a partial measurement of the charge, with a resulting spin wavefunction).  We will denote this quantity as $S_A^{s/c}$.  If an eigenstate $\ket{\psi}$ is in an area law after the partial measurement of either the spin or the charge on each site, then we refer to $\ket\psi$ as a ``quantum disentangled eigenstate.''

Let us now summarize the procedure for performing the diagnostic.  Given a subregion $A$ and a finite energy density eigenstate $\ket\psi$ (which we assume exhibits an overall volume law for the entanglement entropy), the QDL diagnostic is as follows. (i) Perform a partial measurement of the system, by measuring the spin on each site, which gives some spin configuration $\{\sigma\}$.  (ii) Consider the post-measurement wavefunction, $\ket{\phi_{\{\sigma\}}}$. (iii) Calculate the post-measurement entanglement entropy, $S_A(\ket{\phi_{\{\sigma\}}})$.  (iv) Average this quantity over all possible measurement outcomes, weighted by their Born rule probabilities, to obtain $S_A^{c/s}$. (v) Consider the scaling of $S_A^{c/s}$ with subsystem size $L_A^d$ to identify whether it scales with the boundary size or the volume of region $A$.  If it scales with the boundary, then $\ket\psi$ is a quantum disentangled eigenstate.

The partial measurements considered can be implemented in experiments on cold atomic gases, and it has recently become possible to measure the \Renyi entanglement entropy (a close cousin of the von Neumann entropy) in cold atomic systems \cite{Alves2004-entanglement, Daley2012-measuring-entanglement, Islam2015-renyi-measurement, Kaufman2016x-thermalization-expt}.  We will further discuss these connections in \Sec{sec:discussion}.

Having introduced the entanglement entropy diagnostic for quantum disentangled eigenstates, we now turn to the model on which we will focus for the remainder of the paper.

\begin{figure*}[htb]
  \centering
    \sidesubfloat[]{\includegraphics[width=0.45\textwidth]{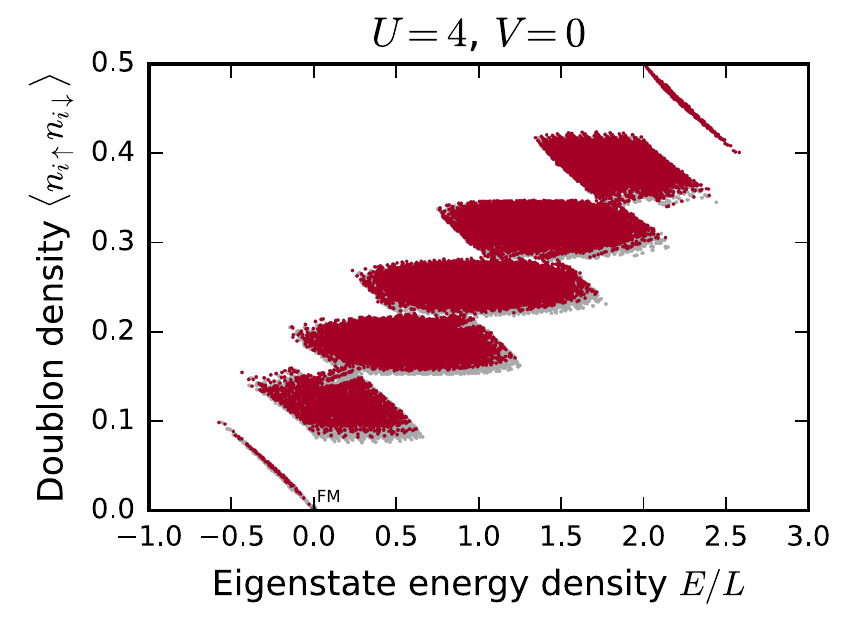}\label{subfig:doublon_hubbard_largeU}}
    \sidesubfloat[]{\includegraphics[width=0.45\textwidth]{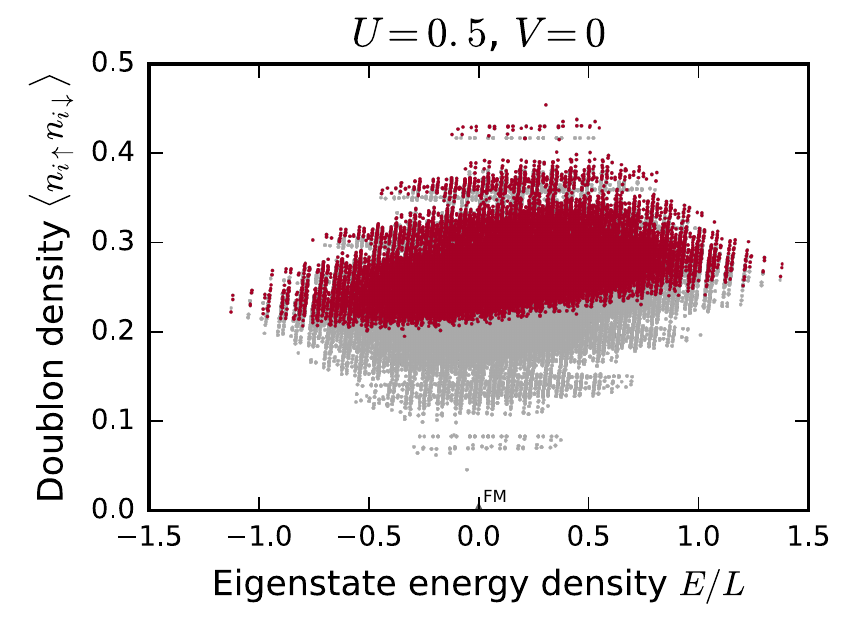}\label{subfig:doublon_hubbard_smallU}} \\
    \sidesubfloat[]{\includegraphics[width=0.45\textwidth]{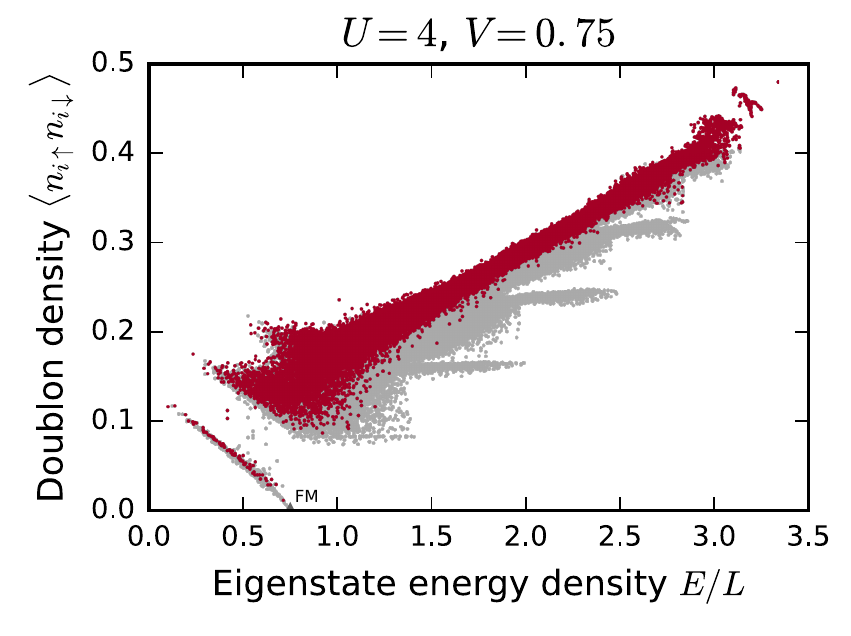}\label{subfig:doublon_V_largeU}}
    \sidesubfloat[]{\includegraphics[width=0.45\textwidth]{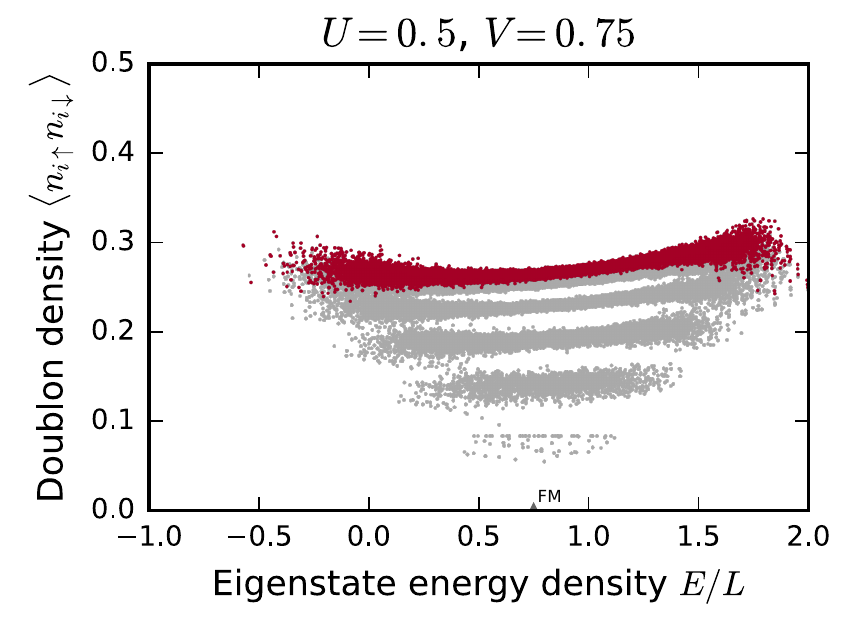}\label{subfig:doublon_V_smallU}}
  \caption{Doublon occupancy of the itinerant electron model (\Eq{eq:model}) obtained from exact diagonalization at system size $L=12$.  Plotted are all eigenstates at half-filling ($N_\up = N_\dn = L/2$), with total spin singlets emphasized in red.  The top panels show the pure Hubbard model ($V=0$), while the bottom panels show the model with an additional nearest-neighbor repulsion term ($V=3/4$), which breaks integrability.  The left panels show a ``large'' value of $U$, while the right panels show results for ``small'' $U$.  In each plot, the Heisenberg ferromagnet (the unique state with maximum total spin) is plotted in dark grey and labeled ``FM.''}
  \label{fig:doublon_results}
\end{figure*}

\section{Model} \label{sec:model}

We consider the 1D Hubbard chain with an additional nearest-neighbor repulsion term:
\begin{align}
\label{eq:model}
H &= H_\mathrm{Hubbard} + H_V \\
H_\mathrm{Hubbard} &= -t\sum_{\langle ij \rangle \, \sigma}\left( c^\dagger_{i\sigma}c_{j\sigma} + \mathrm{H.c.} \right) + U \sum_{i} n_{i\up} n_{i\dn} \nonumber \\
H_V &= V \sum_{\langle ij \rangle} n_i n_j \nonumber
\end{align}
where $n_{i\sigma} = c^\dagger_{i\sigma}c_{i\sigma}$, $n_i = n_{i\uparrow} + n_{i\downarrow}$, and $\sum_{\langle ij \rangle}$ denotes a sum over nearest neighbors.  The spin label $\sigma$ takes the values $\{\up,\dn\}$.  The 1D Hubbard chain is solvable exactly by Bethe ansatz \cite{Lieb1968-hubbard-1d, Essler2005-book} and is therefore not expected to exhibit eigenstate thermalization due to its integrability.  Hence we add the nearest-neighbor repulsion term, which breaks integrability when $V \ne 0$.  We will consider periodic boundary conditions throughout.  We choose an overall energy scale by setting $t=1$.

The model in \Eq{eq:model} has a number of symmetries.  It conserves total particle number $N \equiv N_\up + N_\dn$, where $N_\sigma \equiv \sum_i n_\sigma$.  Momentum is conserved due to translation invariance.  The system also conserves total $S^z \equiv \sum_i S^z_i$ and total spin $\sum_{ij} \mathbf{S}_i \cdot \mathbf{S}_j$.

If the lattice is bipartite (in 1D, if the number of sites $L$ is even), there is an additional symmetry in the Hubbard model, which can be seen by considering a particle-hole transformation on the down spin species: $c_{j\dn} \rightarrow (-)^j c^\dagger_{j\dn}$.  This transformation leaves the kinetic term invariant but maps the $U$ term to $-U$ in $H_\mathrm{Hubbard}$.  It also implements the transformation $n_j \rightarrow \sigma^z_j + 1$ and $\sigma^z_j \rightarrow n_j - 1$, thus mapping the spin sector to charge sector and vice-versa.  Because of this duality, it is apparent that the Hubbard model has a ``hidden'' charge SU(2) symmetry in addition to its spin SU(2) symmetry, resulting in an enlarged symmetry group, SO(4) \cite{Yang1989-eta-pairing, Yang1990-SO4}.  This transformation also maps the nearest-neighbor repulsion term $H_V$ to a nearest-neighbor spin term, $4V \sum_{\langle ij \rangle} S^z_i S^z_j$.  As a result, the $V$ term breaks the charge SU(2) symmetry.

In this paper we will focus on the above Hamiltonian with positive $U$, and consider the entanglement entropy after a partial measurement of the spin.  Because of the above duality transformation, this is equivalent to considering a negative-$U$ Hubbard model with a nearest-neighbor $S^z_i S^z_j$ exchange term, and the entanglement entropy after a partial measurement of the charge degrees of freedom.  Although we will focus on the positive-$U$ model, we will not hesitate to discuss the equivalent physics in the negative-$U$ model when doing so can guide intuition.

\section{Numerical details} \label{sec:numerical}

To investigate the properties of eigenstates of the Hamiltonian (\Eq{eq:model}), we perform numerical exact diagonalization calculations.  When performing exact diagonalization, it is advantageous to represent the Hamiltonian in block-diagonal form, taking advantage of as many symmetries as possible.  This allows one to perform the numerical diagonalization separately in each symmetry sector, each of which has a reduced basis size.  The model conserves both spin-up and spin-down particle number separately.  We focus on half filling ($N_\up = N_\dn = L/2$), in which case the model also has spin-flip and particle-hole symmetries.  Due to periodic boundary conditions, the model also conserves momentum, allowing the physics to be considered in each momentum sector independently.  We exploit each of these abelian symmetries.

The non-abelian SU(2) spin symmetry of the model leads to additional conserved quantities.  Because it is much more difficult to take advantage of non-abelian symmetries in exact diagonalization, we explicitly break the degeneracy due to the SU(2) spin symmetry by adding a total spin term $\sum_{ij} \mathbf{S}_i \cdot \mathbf{S}_j$ to the Hamiltonian with large, irrational coefficient.  This does not change the physics in any given sector, but does allow us to obtain eigenstates of the Hamiltonian that are also eigenstates of the SU(2) total spin operator.  As discussed in \Sec{sec:model}, the pure Hubbard model ($V=0$) on a bipartite lattice has a second SU(2) ``pseudo-spin'' symmetry, which is due to the symmetry between the charge and spin sectors.  For this reason, we also add a total pseudo-spin term to the model when $V=0$ to break the degeneracies arising from this symmetry.

At system sizes where computational resources permit, we perform a full diagonalization of the system in each momentum sector independently.  For larger system sizes, we use ARPACK \cite{ARPACK} to obtain several hundred eigenvalue/eigenvector pairs that are lowest in energy.  In each case, we study the system at half filling and focus on total spin singlets.

\section{Doublon expectation value results} \label{sec:doublon_results}

In this section we examine the expectation value of the doublon density $\< n_{i\up} n_{i\dn} \>$ for each eigenstate in the many-body spectrum.  (Because the system is translationally invariant, this quantity is independent of site $i$.)

\subsection{Large $U$}

Let us begin by considering each eigenstate of the large-$U$ Hubbard model ($V=0$), as shown in \Fig{subfig:doublon_hubbard_largeU}.  As mentioned in \Sec{sec:model}, the highest excited state of this model is the ground state of the model with $U \rightarrow -U$, due to the duality resulting from the particle-hole transformation on the down spin species.  This symmetry is apparent in the plot, as it is symmetric under a combined horizontal and vertical reflection.  (Note that under this duality, total spin singlets are mapped to states with total pseudo-spin zero, which need not be spin singlets.)

\begin{figure}[tb]
  \centering
  \sidesubfloat[]{\includegraphics[width=0.85\textwidth]{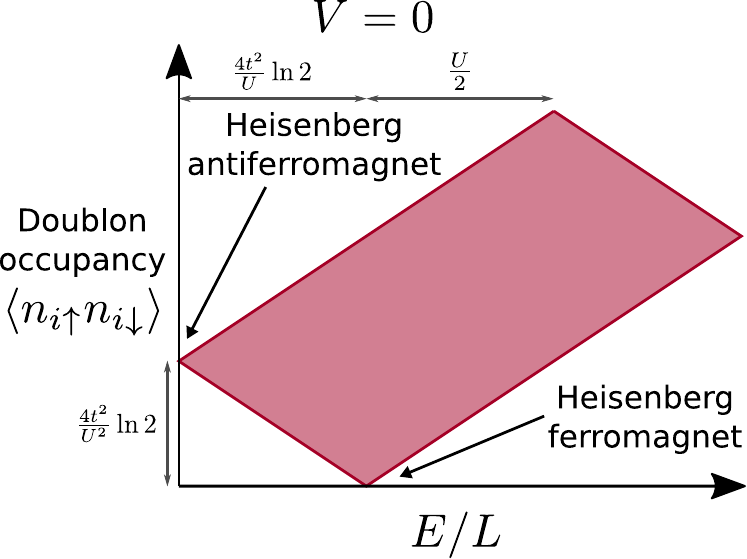}\label{subfig:doublon_sketch_hubbard_largeU}} \\
  \vspace{0.75cm}
  \sidesubfloat[]{\includegraphics[width=0.85\textwidth]{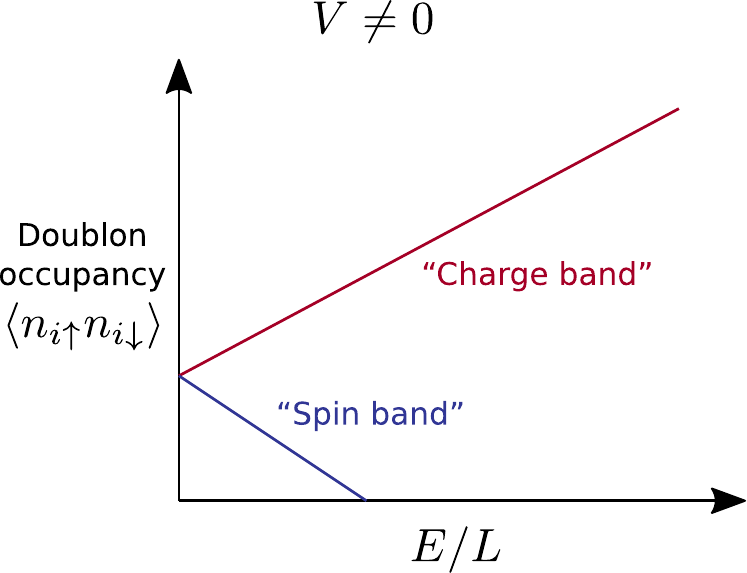}\label{subfig:doublon_sketch_V_largeU}}
  \caption{Putative sketch of the doublon occupancy versus eigenstate energy density for large $U$ in the thermodynamic limit, for both (a) the Hubbard model ($V=0$) and (b) the non-integrable model ($V \ne 0$).  In each case, eigenstates at half-filling which are total spin singlets are considered.}
  \label{fig:doublon_sketch_large_U}
\end{figure}

\begin{figure*}[htb]
  \centering
    \sidesubfloat[]{\includegraphics[width=0.45\textwidth]{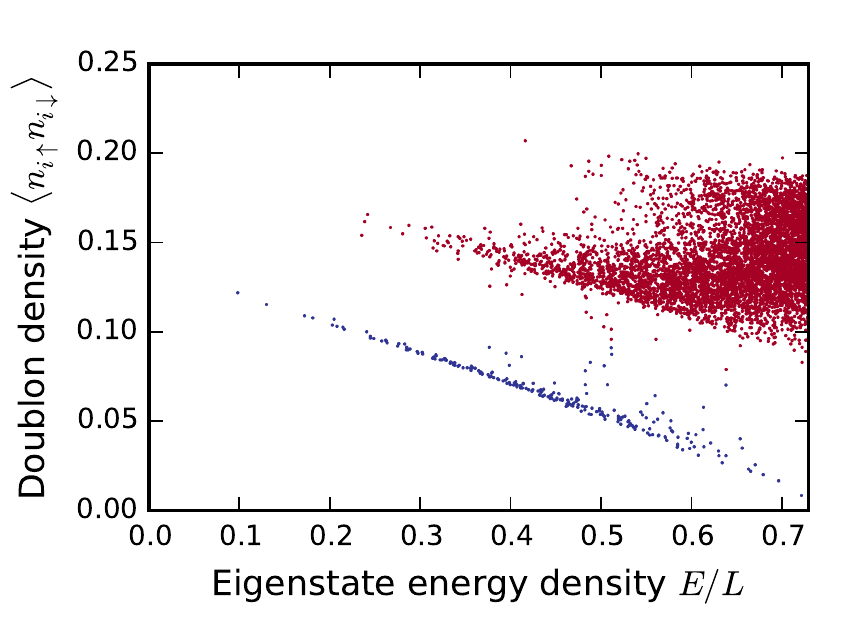}\label{subfig:doublon_L14}}
    \sidesubfloat[]{\includegraphics[width=0.45\textwidth]{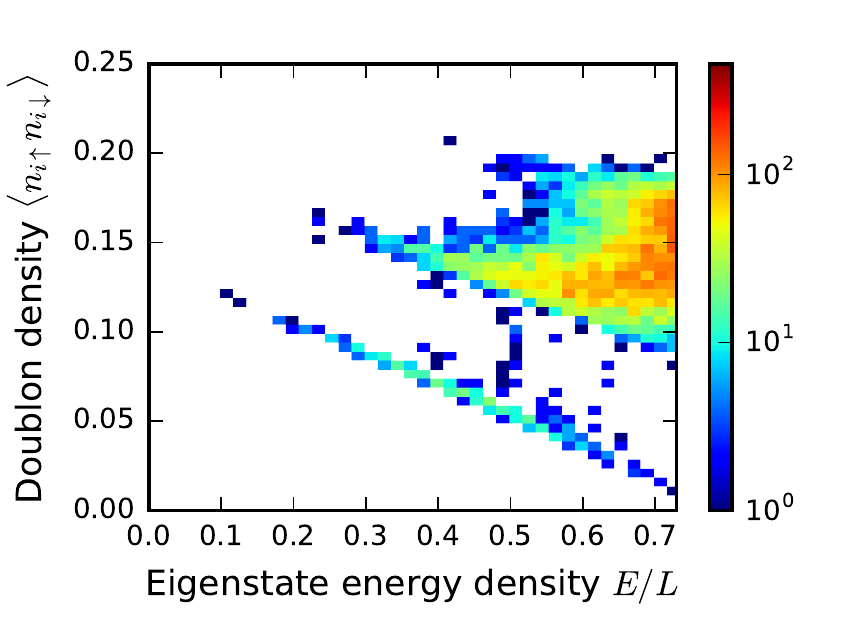}\label{subfig:doublon_L14_hist}}
  \caption{(a) Doublon occupancy for $U=4$, $V=3/4$ at system size $L=14$, as calculated using ARPACK's iterative eigensolver, which returns a portion of the spectrum.  Included are all eigenstates that are total spin singlets at half filling.  The ``spin band'' states are emphasized in blue.  States in all momentum, spin-flip, and particle-hole sectors are combined in this plot.  (b) Logarithmic histogram plot of same quantity.}
  \label{fig:doublon_results_L14}
\end{figure*}

At a given finite energy density in the pure Hubbard model, there exist eigenstates with a range of doublon expectation values.  This is expected for an integrable model, as the plot of a generic expectation value with respect to energy density should fill an area in the thermodynamic limit.  (By contrast, a system which obeys ETH must take a unique expectation value at each energy density.)  In \Fig{subfig:doublon_hubbard_largeU}, it is clear that the results are for a finite size system, as one can easily recognize the bands due to each overall possible doublon count (from $0$ to $L/2$), each offset in energy density by approximately $U/L$.  The band lowest in energy density includes states with low charge fluctuations, the spectrum of which is governed entirely by spin excitations.  In fact, there are $\binom{L}{L/2}$ states in this ``spin band,'' each of which maps to a state in the Heisenberg model restricted to $S_z=0$.  In the thermodynamic limit, the bands will become indistinguishable, resulting in the eigenstates filling a large area of the plot in the shape of a parallelogram.  A sketch of this plot in the thermodynamic limit is given in \Fig{subfig:doublon_sketch_hubbard_largeU}.

The plot's area is bordered on the bottom left by the states in the Heisenberg spin band.  These states can be identified by performing a canonical transformation in powers of $t/U$ \cite{MacDonald1988-expansion, Delannoy2005-hubbard}, resulting in the effective spin-only Hamiltonian
\begin{equation}
H^{(2)}_\mathrm{spin} = \frac{4t^2}{U} \sum_{\<ij\>} \( \mathbf{S}_i \cdot \mathbf{S}_j - \frac14 \),
\end{equation}
which is equivalent to the Heisenberg model with $J = 4 t^2 / U$.  From Bethe ansatz \cite{Bethe1931-ansatz}, the 1D ground state is known to have $\braket{\mathbf{S}_i \cdot \mathbf{S}_{i+1}} = \frac14 - \ln 2$; thus, the ground state energy density is $-\(4t^2/U\) \ln 2$ up to corrections of order $t^3/U^2$.  The ground state doublon expectation value in the anti-ferromagnetic ground state can also be determined to be
\begin{align}
\braket{n_{i\up} n_{i\dn}} &= \frac{1}{L} \frac{\partial}{\partial U} \braket{H^{(2)}_\mathrm{spin}} \nonumber \\
&= 4 \ln 2 \(\frac{t}{U}\)^2,
\end{align}
up to corrections of order $(t/U)^3$.

The Heisenberg ferromagnet, which consists in the $S_z=0$ sector of all spins pointing in the $x$-direction, has doublon expectation value and energy density of precisely zero.  The Heisenberg ferromagnet is itself not a singlet, but it is clear from \Fig{subfig:doublon_hubbard_largeU} that there are overall spin singlet states arbitrarily close to this point.  Note that under the spin-charge duality introduced in \Sec{sec:model}, the Heisenberg ferromagnet maps to the ``$\eta$-paired'' state (first introduced in \Ref{Yang1989-eta-pairing}), which itself has doublon occupancy $\frac12$.

Let us now break integrability by setting $V=3/4$.  Here, common wisdom dictates that full thermalization ought to occur, since the system is non-integrable and contains no disorder.  \Fig{subfig:doublon_V_largeU} shows the doublon expectation value results for large $U$ in this model.  Remarkably, the ``spin band'' of states corresponding to the Heisenberg model remains distinct from the remaining states (which we dub the ``charge band''), even though they overlap in energy density.  In this range of energy densities, the doublon expectation value takes two distinct values, an apparent violation of ETH\@.  The highest excited state in the spin band is the Heisenberg ferromagnet, which remains an eigenstate when $V \ne 0$.  The Heisenberg ferromagnet is the unique state with maximum total spin, and it has zero doublon occupancy.  In its vicinity are spin band states with all possible values of total spin.  \Fig{subfig:doublon_sketch_V_largeU} provides a putative sketch of the $V \ne 0$ doublon occupancy plot in the thermodynamic limit.  We will provide evidence in \Sec{sec:entropy_results} that the states in the spin band are \emph{quantum disentangled eigenstates} according to the definition in \Sec{sec:diagnostic}.

The spin band remains intact for all system sizes accessible to our numerics.  \Fig{subfig:doublon_L14} shows the doublon expectation value for $L=14$ calculated using ARPACK's iterative eigensolver in the range of energy densities where the spin and charge bands overlap.  Note that both this figure and \Fig{subfig:doublon_V_largeU} show states between the spin and charge bands which are in the midst of an avoided crossing if one were to vary $U$ slightly; such states are expected at any finite system size.  The ultimate question is whether these bands remain distinct in the thermodynamic limit.  \Fig{subfig:doublon_L14_hist} shows a 2D histogram of the same quantity, plotted on a logarithmic scale.  Although there are many states in the charge band with which the spin band states could mix, the spin band appears to remain robustly distinct from the charge band, thus supporting the claim that this model violates ETH\@.

\begin{figure}[tb]
  \centering
  \sidesubfloat[]{\includegraphics[width=0.85\textwidth]{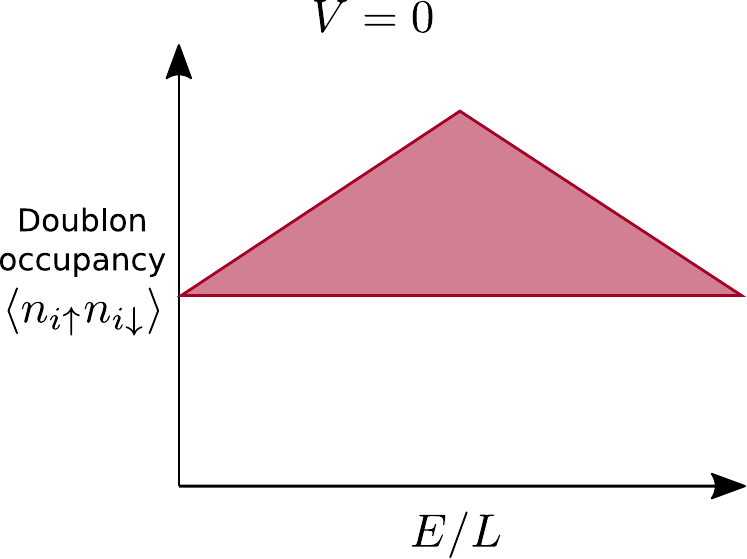}\label{subfig:doublon_sketch_hubbard_smallU}} \\
  \vspace{0.75cm}
  \sidesubfloat[]{\includegraphics[width=0.85\textwidth]{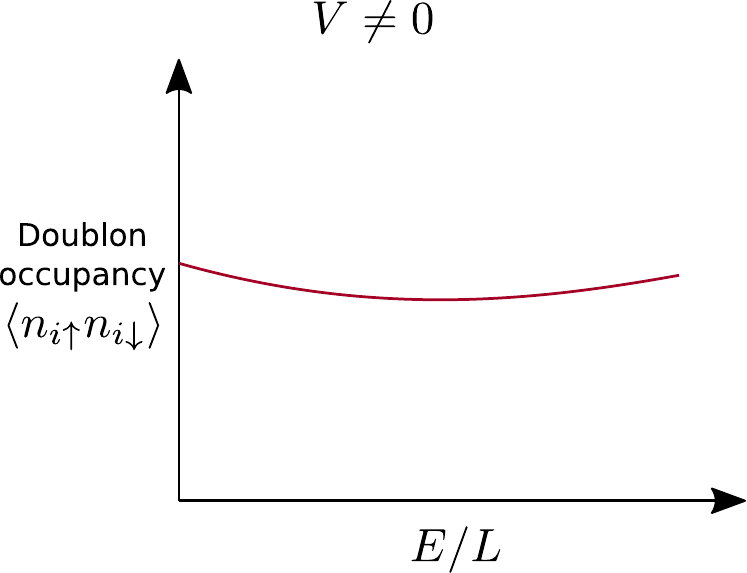}\label{subfig:doublon_sketch_V_smallU}}
  \caption{Putative sketch of the doublon occupancy versus eigenstate energy density for small $U$ in the thermodynamic limit, for both (a) the Hubbard model ($V=0$) and (b) the non-integrable model ($V \ne 0$).  In each case, eigenstates at half-filling which are total spin singlets are considered.}
  \label{fig:doublon_sketch_small_U}
\end{figure}

It is important to note that the regime we are considering in numerics ($U=4$, $V=3/4$, and $t=1$) consists of all parameters of order unity.  Finite size effects are most relevant when the ratio of parameters is of order (or greater than) the total system size \cite{Papic2015x-ti-mbl}.  In the case considered here, the ratio of any two parameters is significantly less than the largest accessible system size, $L=14$.  This suggests that the apparent ETH violation may indeed be robust in the thermodynamic limit.

Counting both singlets and non-singlets, there are $[\binom{L}{L/2}]^2$ total states in the half-filled sector we are considering.  Of these states, $\binom{L}{L/2}$ are in the spin band.  The number of states in the spin band is exponential in system size; however, there are \emph{exponentially more} states in the charge band.  The continued existence of the spin band is therefore a violation of the strongest form of ETH, where non-thermal states vanish in the thermodynamic limit \cite{Biroli2010-rare}.  Such a violation was previously only expected in integrable models.  In principle, the existence of states for which ETH fails implies that there exist initial states with low energy variance that will fail to thermalize at any time\cite{Kim2014-all-eigenstates}.

\begin{figure}[b]
  \centering
    \sidesubfloat[]{\includegraphics[width=0.85\textwidth]{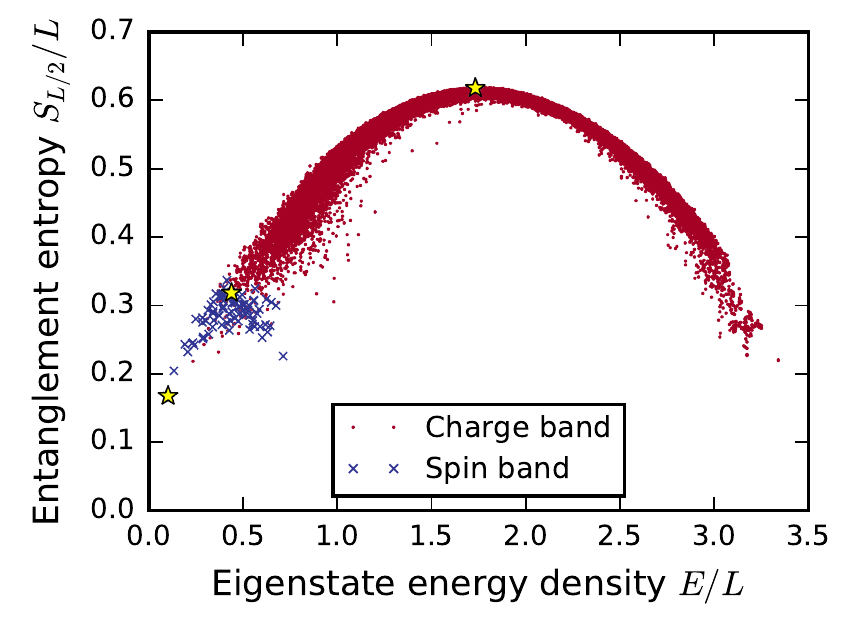}\label{subfig:entropystar}} \\
    \sidesubfloat[]{\includegraphics[width=0.85\textwidth]{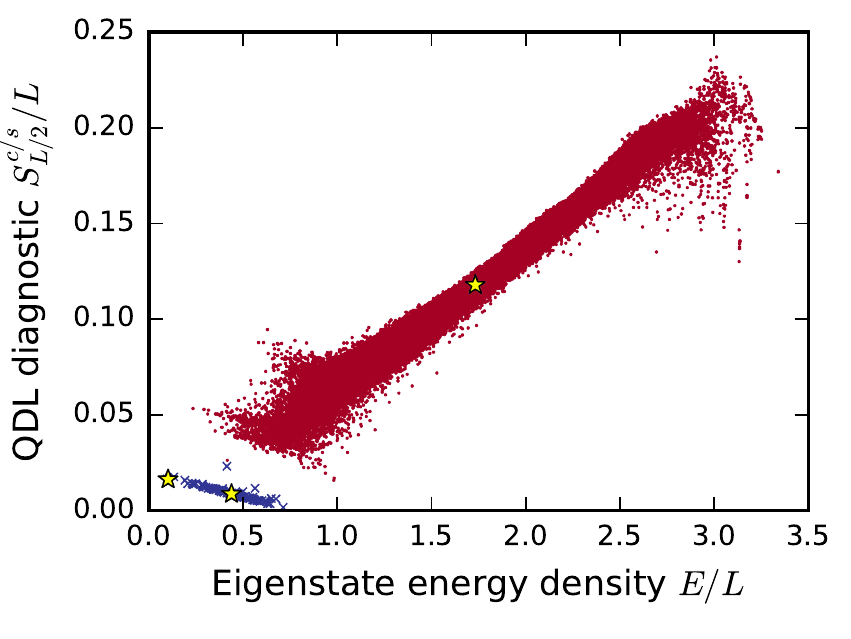}\label{subfig:diagnosticstar}}
  \caption{Numerical half-cut (a) entanglement entropy density $S_{L/2} / L$ and (b) QDL diagnostic density $S_{L/2}^{c/s} / L$ for the model in \Eq{eq:model} at $L=12$, $U=4$, $V=3/4$, and $t=1$, the same non-integrable parameters as Figs.~\ref{subfig:doublon_V_largeU} and \ref{fig:doublon_results_L14}.  Here, all eigenstates that are total spin singlets are plotted, and the states identified to be in the spin band are colored in blue, while charge band states are in red.  The QDL diagnostic shown is the average entanglement entropy after a partial measurement of the spin on each site, as detailed in \Sec{sec:diagnostic}.  Starred are three states that are explored in detail in \Fig{fig:ninepanel}.}
  \label{fig:starred}
\end{figure}

\subsection{Transition to small $U$}

Let us now turn to the physics for small $U$, as shown in the right panels of \Fig{fig:doublon_results}.  We start with the pure Hubbard model (\Fig{subfig:doublon_hubbard_smallU}).  As expected for an integrable model, the eigenstates in this plot fill an area in the doublon--energy-density plane.  One particularly striking feature of this plot is that there no longer exist total spin singlet states which are arbitrarily close to the Heisenberg ferromagnet.  As one decreases $U$, the singlet states appear to ``lift off'' the $x$-axis around $U/t \simeq 1$, regardless of system size.  A proposed sketch of the resulting plot for singlets is shown in \Fig{subfig:doublon_sketch_hubbard_smallU}.  It is an interesting open question whether there exists a critical $U_c$, below which there are no longer singlet states arbitrarily close to the Heisenberg ferromagnet.  The question of whether such an eigenstate phase transition exists in the Hubbard model is expected to be analytically tractable using Bethe ansatz, and we leave this for future work.

\Fig{subfig:doublon_V_smallU} shows the doublon expectation value results for small $U$ in the non-integrable model ($V \ne 0$).  In this parameter regime, the model exhibits strong ETH, although each total spin sector thermalizes to a different value.  \Fig{subfig:doublon_sketch_V_smallU} sketches the expected shape of this plot for singlets only in the thermodynamic limit.

\begin{figure}[tb]
  \centering
  \sidesubfloat[]{\includegraphics[width=0.85\textwidth]{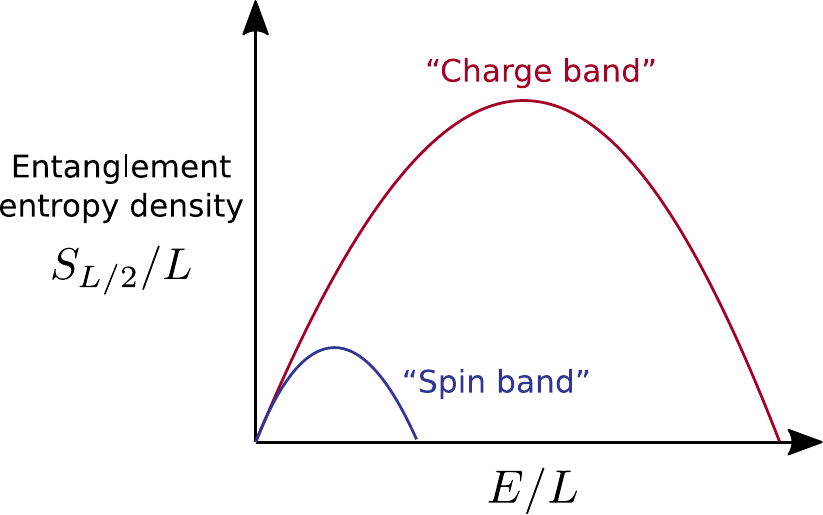}\label{subfig:entropy_sketch}} \\
  \sidesubfloat[]{\includegraphics[width=0.85\textwidth]{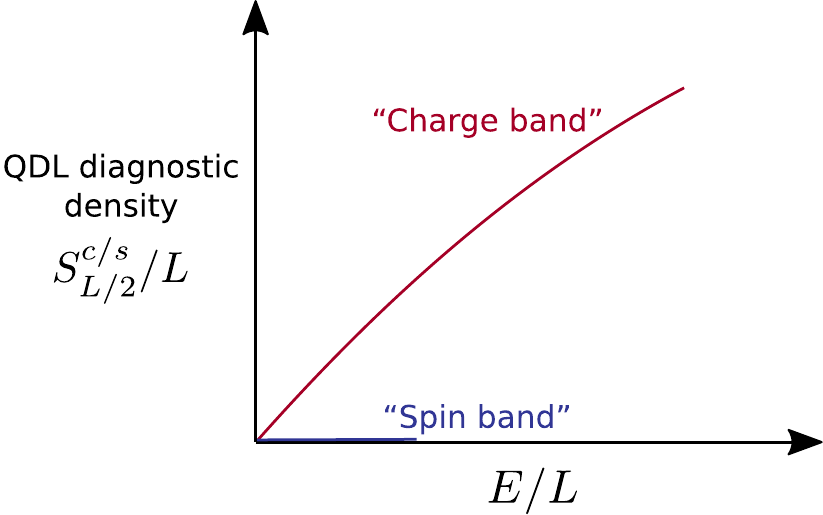}\label{subfig:diagnostic_sketch}}
  \caption{Proposed sketches of (a) the entanglement entropy density and (b) the QDL diagnostic density in the thermodynamic limit, for singlets in the large-$U$ non-integrable model, as based on Figs.~\ref{fig:starred} and \ref{fig:ninepanel}.}
  \label{fig:entropy_sketches}
\end{figure}

\section{Entanglement entropy diagnostic results} \label{sec:entropy_results}

Now that we have provided numerical evidence for the existence of two bands (a ``charge band'' and ``spin band'') in the large-$U$ limit of the non-integrable model (as sketched in \Fig{subfig:doublon_sketch_V_largeU}), we turn toward considering the entanglement entropy and QDL diagnostics, as introduced in \Sec{sec:diagnostic}.

\begin{figure}[b]
  \centering
  \includegraphics[width=\textwidth]{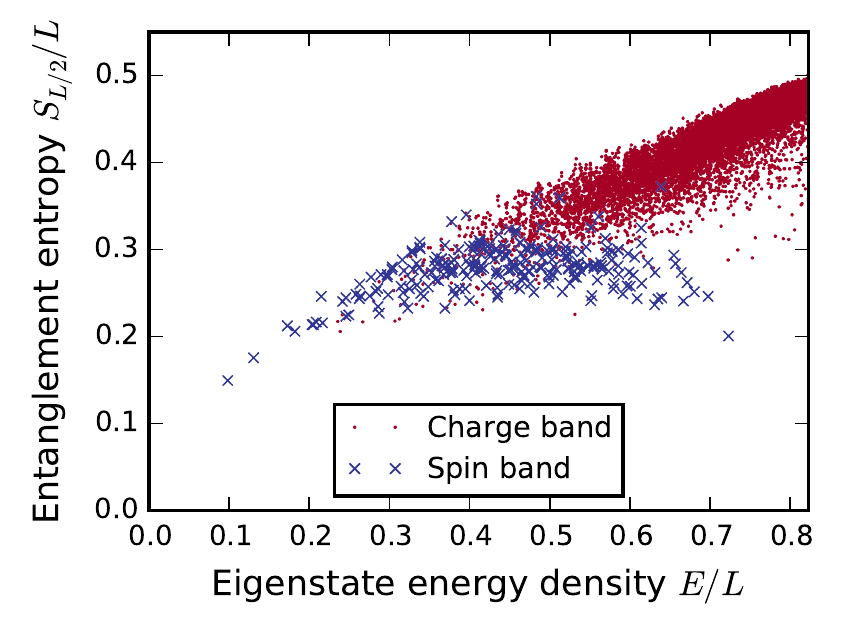}
  \caption{Numerical half-cut entanglement entropy density results for system size $L=14$ at the non-integrable point $U=4$, $V=3/4$.  The ``spin band'' states, as identified in \Fig{subfig:doublon_L14}, are plotted in blue.}
  \label{fig:entropy_L14}
\end{figure}

\Fig{subfig:entropystar} plots the half-cut entanglement entropy density for each eigenstate that is a total spin singlet, with respect to its energy density.  The states identified from \Fig{subfig:doublon_V_largeU} to be in the spin band are colored in blue, while the remaining charge band states are in red.  It appears from this plot that the spin and charge bands form two distinct entropy curves, which overlap in energy density.  In both cases, the entanglement entropy scales linearly with total system size for states with finite energy density, although the states in the spin band have a smaller volume-law coefficient.  \Fig{subfig:entropy_sketch} provides a proposed sketch of this plot in the thermodynamic limit.  Results at $L=14$ further support the existence of two overlapping entropy curves (see \Fig{fig:entropy_L14}).

\begin{figure*}[tb]
  \centering
  \includegraphics[width=\textwidth]{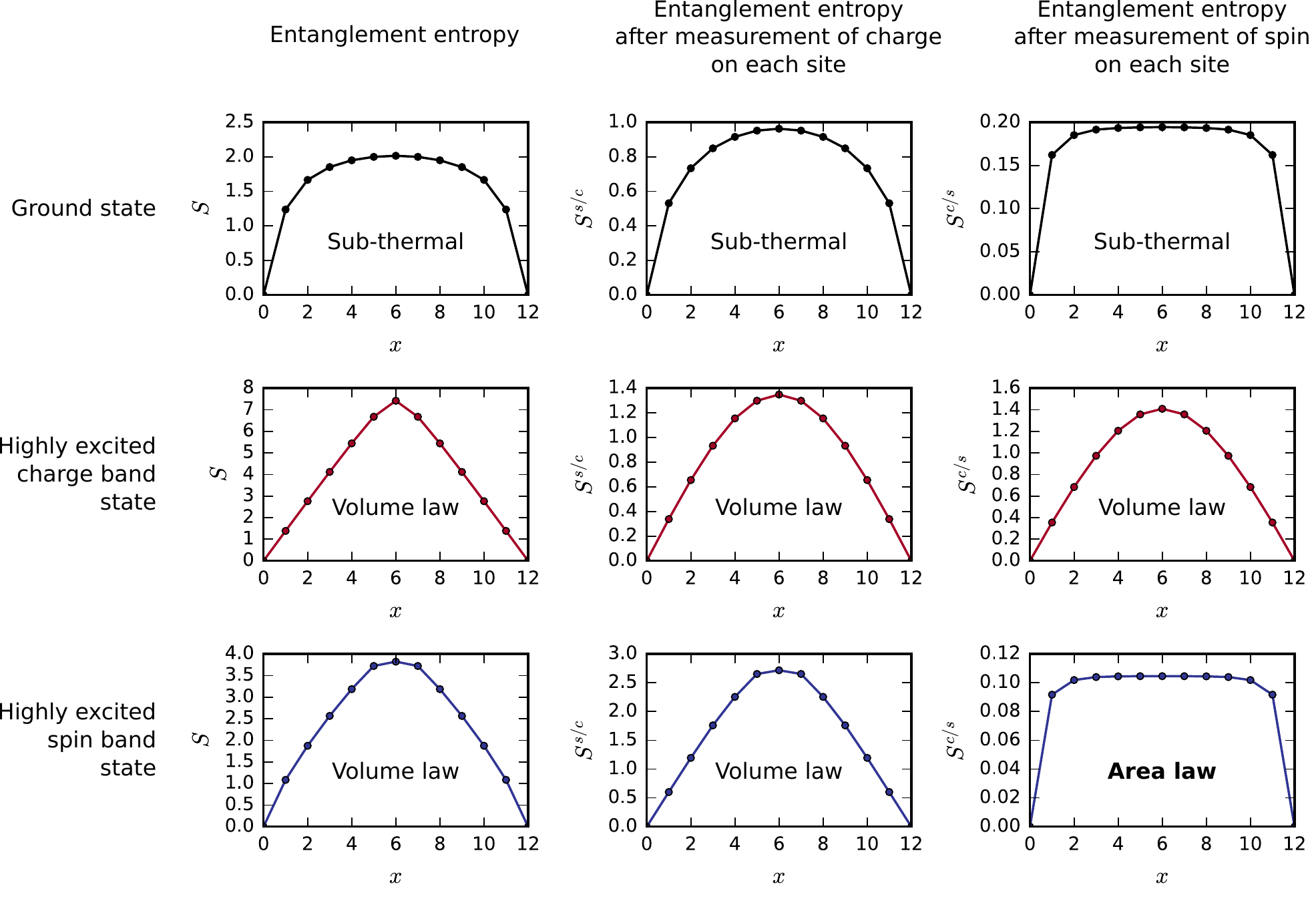}
  \caption{Scaling of the entanglement entropy and QDL diagnostics with subsystem cut size $x$, for the three starred states in \Fig{fig:starred}.  The left column is the overall entanglement entropy.  The middle and right columns plot the QDL diagnostic entanglement entropy after a partial measurement of the charge and spin on each site, respectively.  The top row (in black) shows each quantity plotted for the ground state, each of which scales sub-thermally.  The middle row (in red) shows the quantities for a highly excited state in the charge band, each of which appears to scale as a volume law.  Finally, the bottom row (in blue) shows the three quantities for a highly excited state in the spin band.  Here, $S$ and $S^{s/c}$ both scale as a volume law, but the entanglement entropy after a spin measurement $S^{c/s}$ scales as an area law, thus fulfilling the criteria for a \emph{quantum disentangled eigenstate} as defined in \Sec{sec:diagnostic}.}
  \label{fig:ninepanel}
\end{figure*}

The apparent existence of two distinct, overlapping entropy curves calls into question the basic tenets of quantum statistical mechanics.  Within the context of ETH, entanglement entropy is equal to thermal entropy, and it is possible to assign a ``temperature'' to an eigenstate by identifying $1/T$ to be the slope of the energy-entropy curve\footnote{In general, defining temperature this way will be equivalent to the temperature obtained by matching an eigenstate's energy density to the system's energy density in the canonical ensemble at some temperature.}.  Thus, all states where the entropy has a positive slope are at positive temperatures, the states with maximum entropy are at infinite temperature, and the states where the entropy slope is negative are at negative temperatures.  If we assume \Fig{subfig:entropy_sketch} is correct in the thermodynamic limit, it implies that there are energy densities that contain ``hot'' spin-band states alongside much cooler charge-band states.  If these states are indeed robust as $L \rightarrow \infty$, an isolated quantum system governed by this model will not thermalize according to canonical statistical mechanics.

Let us now consider the QDL diagnostic after a partial measurement of the spin on each site, the half-cut of which is shown in \Fig{subfig:diagnosticstar}.  The spin-band states have greatly reduced entropy after such a partial measurement, as knowledge of the spin state provides nearly all information in these states with very little charge fluctuation.  To further explore the entropy and QDL diagnostic properties of this system, we focus in detail on three states: (i) the ground state; (ii) a highly excited state in the charge band; and (iii) a highly excited state in the spin band.  These three states are represented by stars in \Fig{fig:starred}, and are explored in detail in \Fig{fig:ninepanel}.  Plotted in this figure is the scaling of the entanglement entropy of each state, as well as the scaling of the QDL diagnostics after a partial measurement of the charge or spin on each site.

The scaling properties of the ground state are plotted in the top row of \Fig{fig:ninepanel} (in black).  Because the model is gapless, the ground state's entanglement entropy scales as $\log x$ \cite{CalabreseCardy2004-ee}.  As expected, this sub-thermal entanglement scaling remains after a partial measurement of the charge or spin on each site, as can be seen in the center and right panels of the top row.

The middle row (in red) shows the scaling properties of a high entropy excited state from the charge band.  As expected, the overall entanglement entropy of this state scales extensively with system size, which is consistent with the state being in a volume law.  It remains in a volume law after a partial measurement of the charge or spin on each site (middle and right panels).  As such, this highly excited state in the charge band appears to be fully ergodic.

In the bottom row, we examine the scaling properties of a high entropy state from the spin band (in blue).  As can be seen in the left panel, the overall entanglement entropy of this state scales as a volume law, as is expected for a state with finite energy density.  The middle panel considers the QDL diagnostic $S^{s/c}$ which measures the entropy remaining after a partial measurement of the charge on each site.  Because spin-band states have little charge fluctuations, such a measurement obtains very little information about the state, and the post-measurement state is still in a highly-entangled, volume law state.  The bottom right panel of \Fig{fig:ninepanel} shows the QDL diagnostic $S^{c/s}$, the entanglement entropy after a partial measurement of the spin on each site.  Remarkably, this plot saturates to a constant and scales as an \emph{area law}, thus fulfilling the criteria of a quantum disentangled liquid.  The partial measurement of the spin degrees of freedom \emph{disentangles} the charge degrees of freedom in the state, transforming the wavefunction from a volume law to an area law.  This result is consistent with the states breaking ergodicity.

Having established that the states in the spin band are in an area law for the QDL diagnostic $S^{c/s}$, we can form a sketch of the half-cut QDL diagnostic density, which is provided in \Fig{subfig:diagnostic_sketch}.  The QDL diagnostic for the charge band states scales extensively with system size, so this quantity takes a finite value at each finite energy density in the thermodynamic limit.  On the other hand, the spin band states have vanishing QDL diagnostic density in the thermodynamic limit since $S^{c/s}$ scales only with the size of the \emph{boundary} between subregions.

The QDL diagnostic thus acts as a tool for identifying the breakdown of full thermalization.  It provides a qualitative distinction between states in the charge band and those in the spin band---in other words, volume law states which are fully thermal and those which are not.

\section{Discussion} \label{sec:discussion}

In this paper, we have provided numerical evidence for the violation of ETH in a non-integrable system without disorder.  The model, given by \Eq{eq:model}, supports two qualitatively distinct bands of eigenstates which overlap in energy density, thus calling into question the general validity of quantum statistical mechanics in translationally invariant systems.

While the model has exponentially many ``spin band'' states, they are nonetheless exponentially rare compared with the more common ``charge band'' states.  This is reminiscent of an integrable system, where ETH is satisfied for all but a vanishing fraction of eigenstates \cite{Biroli2010-rare, Alba2015-eth}.  In both cases, the existence of non-thermal eigenstates implies that there exist initial states with low energy variance that will fail to thermalize.
In principle, any initial state that has non-vanishing overlap with the spin band will never reach thermal equilibrium.  It will be interesting to identify experimentally preparable states that fall in this class.  Could an initial product state---for instance with one fermion of arbitrary spin per site---be sufficient in demonstrating the failure of thermalization?  Other initial states to consider include quenched states, or states that result from adding a finite density of spin excitations to the quantum ground state of \Eq{eq:model}.

Once non-thermalizing initial states have been identified, it will be fascinating to study the system's time evolution from these states numerically.  What observables fail to relax at long times?  Can this provide any additional clues to the mechanism behind the breakdown of ETH?  It would also be particularly interesting to attempt to realize a quantum disentangled liquid experimentally by implementing the model in a cold atomic gas of fermions, similar to recent experiments on many-body localization \cite{DeMarco2013-mbl-expt, Bloch2015x-mbl-expt, Choi2016x-2d-mbl}.  While a nearest-neighbor repulsion term is beyond the reach of current technology, an alternative method would involve realizing the $S^z S^z$ term in the equivalent dual model, which was discussed in \Sec{sec:model}.  In any case, an experiment in an optical lattice should allow access to much larger system sizes than can be simulated numerically.

The definitive distinguishing feature of the putative QDL phase is the area law scaling of the entanglement entropy after a partial measurement of the spin on each site, as introduced in \Sec{sec:diagnostic}.  Remarkably, a recent experiment has measured the \Renyi entanglement entropy $S_2$ in a cold atomic gas of bosons by performing controlled interference between identical copies of the system \cite{Alves2004-entanglement, Daley2012-measuring-entanglement, Islam2015-renyi-measurement, Kaufman2016x-thermalization-expt}.  In principle, it is also possible to measure \Renyi entanglement entropies in cold fermionic gases \cite{Pichler2013-measure-fermion-entropy, Parsons2016x-microscope, Boll2016x-microscope}.  Suppose we have a reliable experimental protocol for preparing a state which overlaps fully with states in the spin band.  We could then identically prepare two copies of the system and perform a partial measurement on each.  Unfortunately, it is very unlikely that the two copies would exhibit the same measurement outcome, and it follows that the quantum states of the two systems will almost certainly be different.  Because the \Renyi measurement protocol relies on identical copies of a state, it thus cannot be implemented after a partial measurement.  In the end, measuring the average post-measurement entanglement entropy may require performing full quantum tomography on the state resulting after each possible measurement outcome, which is a daunting task.  Let us emphasize that while this diagnostic is unlikely to be implemented in experiment, the mere demonstration of the breakdown of thermalization is likely to be a much easier task.  Along these lines, existing experiments on realizing MBL phases have focused on observables that fail to thermalize, not on demonstrating the area-law scaling of entanglement entropy for eigenstates.

It is worth considering what role symmetries play in the breakdown of ETH in a translationally invariant system.  In this paper we considered the itinerant fermion model only at half filling, but it would be interesting to investigate whether QDL states exist at other filling fractions as well.  Likewise, to what degree is the observed ETH violation \emph{dependent} on symmetries?  The spin band states only exist in certain sectors of total spin, particle-hole parity, and spin-flip parity.  What is special about these sectors which harbor QDL behavior?  Interestingly, breaking \emph{both} the charge and spin SU(2) symmetries seems to eliminate the spin band.  One is tempted to wonder: is a non-abelian symmetry necessary for realizing QDL behavior?

On the other hand, with so many symmetries one must be wary of finite size effects, as each sector contains fewer states with which to mix.  In \Ref{Kim2014-all-eigenstates} it was found that sectors with additional symmetries typically have more pronounced outlier states at a given system size.  Still, each sector we consider has a Hilbert space size comparable to, if not larger than, the best ETH studies to date.  As we have shown above, numerical results up to system sizes of $L=14$ support the existence of the spin band and thus the violation of ETH\@---a clear demonstration of a QDL regime.  If one wishes to establish the QDL as a true phase of matter in this model, the ultimate question, of course, is whether the spin band continues to exist in the thermodynamic limit.  One method for determining the fate of the spin band is to examine the level spacing statistics between the spin band and charge band as the system size is increased, similar to studies of MBL \cite{Oganesyan2007-localization}.  Unfortunately, because there is no disorder over which to average, it is very difficult to get good statistics.  Even if one averages over all possible twists of boundary conditions, the energy level spacings are still highly correlated with each other among samples.  An idea worth investigating is to consider a range of values for the parameters $U$ and $V$, in addition to all possible twists of boundary conditions.

The spin band states exist only in the large-$U$ limit of \Eq{eq:model}, and another interesting task would involve constructing a canonical transformation in powers of $t/U$, transforming Heisenberg eigenstates into eigenstates of \Eq{eq:model} in the spirit of Refs.~\onlinecite{MacDonald1988-expansion} and \onlinecite{Delannoy2005-hubbard}.  This would in principle allow access to larger system sizes, and such a transformation may provide insight into (or a technique for perturbatively proving) the breakdown of thermalization.  For recent work in this direction, see \Ref{Veness2016x-qdl}.

Finally, it should be emphasized that ETH violation in a translationally invariant system has implications beyond condensed matter physics.  In particular, it was recently argued that ETH is itself analogous to the ``no-hair theorem'' for classical black holes \cite{Khlebnikov2013x-thermalization}.  In other words, the statement of ETH parallels the idea that the metric is completely determined by the energy density of a black hole.  The existence of a featureless model that violates ETH may thus have implications for quantum gravity.

In conclusion, using state-of-the-art numerics we have provided evidence for the violation of ETH in a non-integrable model of itinerant electrons.  Our results suggest that this model realizes two distinct bands of energy eigenstates, which overlap in energy density and can be distinguished by a universal, qualitative diagnostic based on the entanglement entropy after a partial measurement.  Because the number of ETH-violating states scales extensively with the system size, there exist initial states that will never reach thermal equilibrium, thus calling into question the validity of quantum statistical mechanics.

\acknowledgments

We are grateful to Leon Balents, Bela Bauer, Erez Berg, Dominic Else, Fabian Essler, Keith Fratus, Steven Girvin, Tarun Grover, Katharine Hyatt, Robert Konik, Cheng-Ju Lin, Lesik Motrunich, Markus M\"uller, Chetan Nayak, Gil Refael, Sid Parameswaran, Neil Robinson, Mauro Schiulaz, Thomas Veness, and David Weld for enlightening discussions regarding this work.
This research was supported in part by the National Science Foundation, under Grant No.\ DMR-14-04230 (JRG and MPAF), by the Walter Burke Institute for Theoretical Physics at Caltech (RVM), and by the Caltech Institute of Quantum Information and Matter, an NSF Physics Frontiers Center with support of the Gordon and Betty Moore Foundation.
We acknowledge support from the Center for Scientific Computing at the CNSI and MRL: an NSF MRSEC (DMR-1121053) and NSF CNS-0960316.

\bibliography{../../../bibtex-master}

\end{document}